\documentclass{ws-procs9x6}
\usepackage{mathrsfs}
\usepackage{color}

\newcommand{\gp}{$\vec \gamma p \rightarrow p\pi^+\pi^-${ }}

\begin{document}

\title{CLAS: Double-Pion Beam Asymmetry} 

\author{S.~Strauch for the CLAS Collaboration}

\address{University of South Carolina\\
Department of Physics and Astronomy\\
712 Main Street\\
Columbia, SC 29208, USA\\
E-mail: strauch@sc.edu}
 
\maketitle

\abstracts{Beam-helicity asymmetries for the \gp reaction have been
  measured for center-of-mass energies between 1.35 GeV and 2.30 GeV
  at Jefferson Lab with the CEBAF Large Acceptance Spectrometer using
  circularly polarized tagged photons.  The beam-helicity asymmetries
  vary with kinematics and exhibit strong sensitivity to the dynamics
  of the reaction, as demonstrated in the comparison of the data with
  results of various phenomenological model calculations. These models
  currently do not provide an adequate description of the data over
  the entire kinematic range covered in this experiment. Additional
  polarization observables are accessible in an upcoming experiment at
  Jefferson Lab with polarized beam and target.}

\section{Introduction}

The properties of the excited states of baryons reflect the dynamics
and relevant degrees of freedom within them. The study of nucleon
resonances is thus an important avenue to learn about the strong
interaction. To disentangle and study experimentally these many,
mostly weak, and largely overlapping resonances is a challenging task,
especially at higher energies. Many nucleon resonances in the mass
region above 1.6~GeV decay predominantly through either $\pi\Delta$ or
$\rho N$ intermediate states into $\pi\pi N$ final states (see the
Particle-Data Group review\cite{pdg04}). This is the region where
resonances are predicted by symmetric quark models, but have not been
observed in the $\pi N$ channel (the so-called ``missing''
resonances); yet may couple strongly to channels like $\pi
\Delta$.\cite{capstick94} This makes electromagnetic double-pion
production an important tool in the investigation of the structure of
the nucleon and the most promising approach is to employ polarization
degrees of freedom in the measurement. There exist a rather large
amount of unpolarized cross-section data of double-pion photo-
and electroproduction on the proton;\cite{pipiN-cs} the amount of
polarization observables in these reactions, however, remains quite
sparse.\cite{pipiN-pol}

The CLAS Collaboration published recently beam-helicity-asymmetry data
in the \gp reaction for energies $W$ between 1.35~GeV and 2.30~GeV in
the center of mass, where the photon beam is circularly polarized and
neither target nor recoil polarization is specified.\cite{Strauch05}
These novel data combine the study of the double-pion final state with
the sensitivity of polarization observables. The beam-helicity
asymmetry is one of 64 polarization observables in the \gp reaction
and is defined as\cite{Roberts04}
\begin{equation}
   I^\odot = \frac{1}{P_\gamma} \cdot 
   \frac{\sigma^+ - \sigma^-}{\sigma^+ + \sigma^-},
\end{equation}
where $P_\gamma$ is the degree of circular polarization of the photon
and $\sigma^\pm$ are the cross sections for the two photon-helicity
states $\lambda_\gamma=\pm 1$.

In this contribution, we give a brief overview of our data, compare
with results of phenomenological models, demonstrate the sensitivity
of this observable to the dynamics of the reaction, and give an
outlook on further studies of other polarization observables.

\section{Experiment}

\begin{figure}[!ht]
\centerline{\includegraphics[width=0.8\columnwidth]{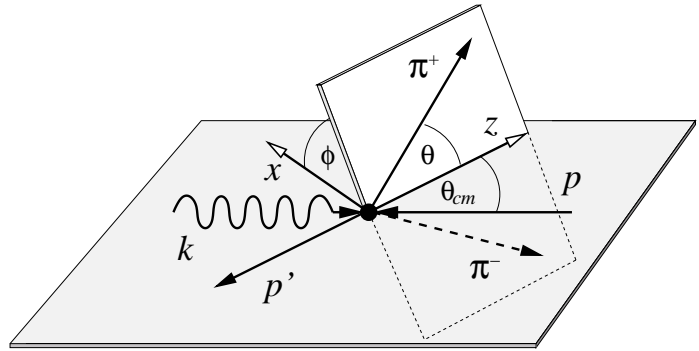}}
\caption{\label{fig:def} Angle definitions for the circularly
polarized real-photon reaction \gp; $\theta_{cm}$ is defined in the
overall center-of-mass frame, and $\theta$ and $\phi$ are defined as
the $\pi^+$ polar and azimuthal angles in the rest frame of the
$\pi^+\pi^-$ system with the $z$ direction along the total
momentum of the $\pi^+\pi^-$ system (helicity frame).}
\end{figure}

The \gp reaction was studied with the CEBAF Large Acceptance
Spectrometer (CLAS)\cite{Mecking03} at Thomas Jefferson National
Accelerator Facility (Jefferson Lab).  A schematic view of the
reaction, together with angle definitions, is shown in
Fig.~\ref{fig:def}.
Longitudinally polarized electrons with an energy of 2.445~GeV were
incident on the thin radiator of the Hall-B Photon
Tagger\cite{Sober00} and produced circularly-polarized tagged photons
in the energy range between 0.5~GeV and 2.3~GeV. The collimated photon
beam irradiated a liquid-hydrogen target. The circular polarization of
the photon beam was determined from the electron-beam polarization and
the ratio of photon and incident electron energy.\cite{Maximon59} The
degree of photon-beam polarization varied from $\approx 0.16$ at the
lowest photon energy up to $\approx 0.66$ at the highest energy.  The
\gp reaction channel was identified in this kinematically complete
experiment by the missing-mass technique, requiring either the
detection of all three final-state particles or the detection of two
out of the three particles in the CLAS detector.

A total of $3 \times 10^7$ $p\pi^+\pi^-$ events were accumulated.
Experimental values of the helicity asymmetry were then obtained as
\begin{equation}
   I^\odot_{\rm exp} = \frac{1}{\rule[2mm]{0mm}{1.5mm}\bar P_\gamma} \cdot
     \frac{Y^+ - Y^-}{Y^+ + Y^-}\;,
\end{equation}
where $Y^\pm$ are the experimental yields, corrected for a small
electron-beam-charge asymmetry. The experimental asymmetries have not
been corrected for the CLAS acceptance to avoid systematic
uncertainties. Instead, the data are compared with event-weighted mean
values of asymmetries from model calculations. These mean values of
asymmetries in a kinematical bin are given by
\begin{equation}
  \bar I^\odot_{\rm model} = \frac{1}{\bar P_\gamma} \cdot 
  \frac{1}{N} \sum_{i=1}^N P_{\gamma,i} I^\odot_i,
\end{equation}
where the sum runs over all $N$ events observed in that bin;
$P_{\gamma,i}$ and $I^\odot_i$ are the degree of circular beam
polarization and the model asymmetry, respectively, for the kinematics
of each of those events. The only major source of systematic
uncertainty is the degree of the beam polarization, which is known to
about 3\%. The uncertainty from the beam-charge asymmetry is
negligible (less than $10^{-3}$).

\section{Results}

\begin{figure}[h]
\centerline{\includegraphics[width=\columnwidth]{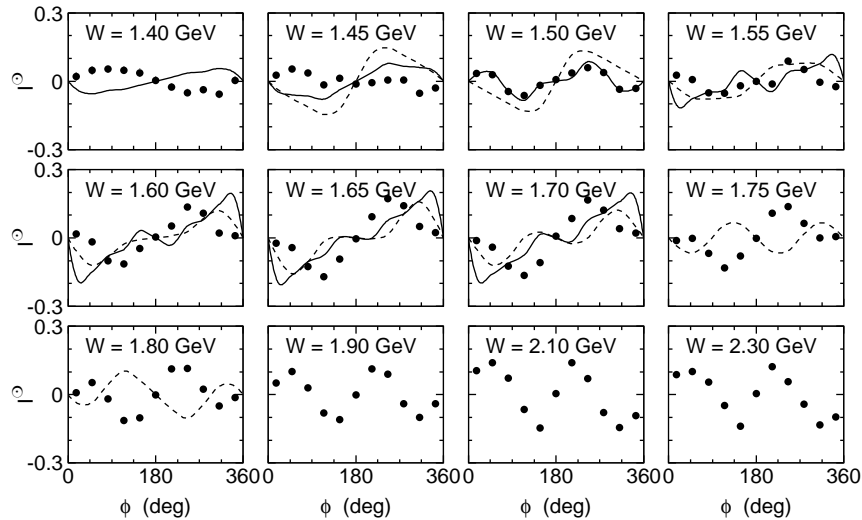}}
\caption{\label{fig:all} Angular distributions for selected
  center-of-mass energy bins (each with $\Delta W = 50$ MeV) of the
  beam-helicity asymmetry for the \gp reaction. The data are
  integrated over the detector acceptance. The statistical
  uncertainties are smaller than the symbol size. The dashed and solid
  curves are the results from model calculations by Mokeev {\it et
    al.}\protect\cite{Mokeev} (for $1.45\; {\rm GeV} \le W \le 1.80\;
  {\rm GeV}$), and by Fix and Arenh{\"o}vel\protect\cite{Fix04} (for
  $W \le 1.70$ GeV), respectively.}
\end{figure}

Figure \ref{fig:all} shows $\phi$-angular distributions of the
helicity asymmetry for various selected 50-MeV-wide center-of-mass
energy bins between $W = 1.40$~GeV and $2.30$~GeV.
The data are integrated over the full CLAS acceptance. The analysis
shows large asymmetries which change markedly with $W$ up to 1.80~GeV;
thereafter they remain rather stable. The asymmetries are odd
functions of $\phi$ and vanish for coplanar kinematics ($\phi=0$ and
$180^\circ$), as expected from parity conservation.\cite{Roberts04}
Thus only sine terms contribute to the Fourier expansion of these
distributions, $\sum a_k \sin (k\phi)$. If two of the final-state
particles are identical, like in the $\pi^0\pi^0p$ final state, the
form of the angular distributions is further restricted and only
even-order terms enter the Fourier series. This has indeed been
observed by the CLAS collaboration in beam-helicity asymmetry data in
the $\vec\gamma ^3{\rm He} \to ppn$ reaction, where the lowest order
term is $a_2\sin(2\phi)$.\cite{Ukwatta05}

The data are compared with results of available phenomenological
models. In the approach by Mokeev {\it et al.}\cite{Mokeev} (dashed
curves), double-charged-pion photo- and electroproduction are
described by a set of quasi-two-body mechanisms with unstable
particles in the intermediate states: $\pi\Delta$, $\rho N$, $\pi
N(1520)$, $\pi N(1680)$, $\pi \Delta(1600)$ and with subsequent decays
to the $\pi^+ \pi^- p$ final state. Residual direct $\pi^+ \pi^- p$
mechanisms are parametrized by exchange diagrams. The model provides a
good description of all available CLAS cross-section and world data on
double-pion photo- and electroproduction at $W<1.9$ GeV and $Q^2<1.5$
GeV$^2$. Results have also been obtained by Fix and Arenh{\"o}vel
using an effective Lagrangian approach with Born and resonance
diagrams at the tree level.\cite{Fix04} The corresponding results are
shown in Fig.~\ref{fig:all} as solid curves.

Although both models previously provided a good description of
unpolarized cross sections, neither of the models give a reasonable
description of the present beam-asymmetry data over the entire
kinematic range covered in this experiment. Even though the model
predictions agree remarkably well for certain conditions, for other
conditions they are much worse and sometimes even out of phase
entirely.

\begin{figure}[!ht]
\centerline{\includegraphics[width=1.0\columnwidth]{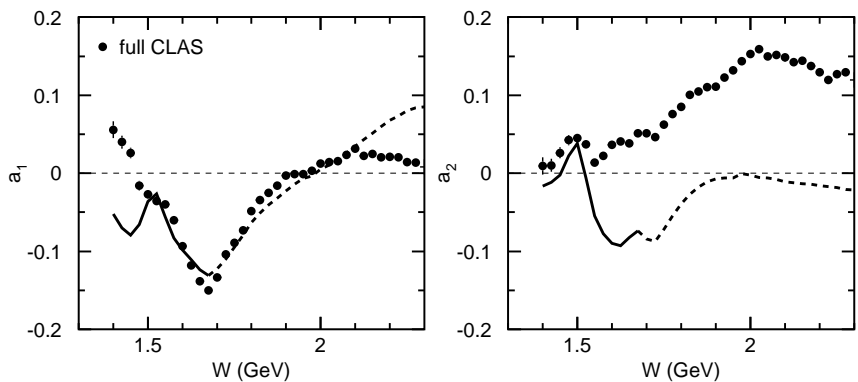}}
\caption{\label{fig:fourier} First ($a_1$) and second ($a_2$) order
  Fourier components of the beam-helicity asymmetry as a function of
  the $\gamma p$ center-of-mass energy.  The data are integrated over
  the full CLAS acceptance and compared with model calculations by Fix
  and Arenh{\"o}vel\protect\cite{Fix04} within (solid) and outside
  (dashed) the range of validity of that model ($W<1.7$~GeV).}
\end{figure}

The result of a Fourier analysis of the $\phi$ distributions is shown
in Fig.~\ref{fig:fourier} and compared with model calculations by Fix
and Arenh{\"o}vel.\cite{Fix04} Apart of the $W = 1.4$~GeV region,
excellent agreement is achieved in these calculations for the $a_1$
coefficient; yet the distribution of the $a_2$ coefficients deviates
from the data above $W \approx 1.55$~GeV. It would be interesting to
find out if there are specific reaction mechanisms contributing to
specific terms in the Fourier decomposition of the angular
distributions.

\begin{figure}[!ht]
\centerline{\includegraphics[width=\columnwidth]{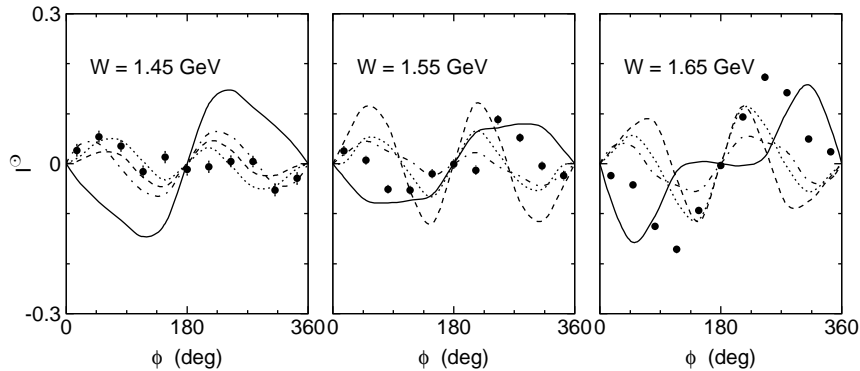}}
\caption{\label{fig:phases} Integrated angular distributions for
  selected center-of-mass energy bins (each with $\Delta W = 50$ MeV)
  of the beam-helicity asymmetry for the \gp reaction.  The solid,
  dashed, dotted, and dash-dotted curves are results from model
  calculations by Mokeev {\it et al.}\protect\cite{Mokeev} with relative phases of
  $0$, $\pi/2$, $\pi$, and $3\pi/2$ between the background- and
  $\pi\Delta$-subchannel amplitudes, respectively.}
\end{figure}

The main theoretical challenge for double-pion photoproduction lies in
the fact that several subprocesses may contribute, even though any
given individual contribution may be small.  In his recent work
Roca\cite{Roca04} studies the beam-helicity asymmetry and our
preliminary data\cite{Strauch04} in the framework of the Valencia
model for double-pion photoproduction. He shows that the shape and
strength of the asymmetries depend strongly on the internal mechanisms
and interferences among different contributions to the process. In
this connection, the polarization measurements should be very helpful
in separating the individual terms.  The particular sensitivity of the
beam asymmetry to interference effects among various amplitudes is
illustrated in Fig.~\ref{fig:phases}.
The solid, dashed, dotted, and dash-dotted curves are the results from
model calculations by Mokeev {\it et al.}\cite{Mokeev} with relative phases of
$0$, $\pi/2$, $\pi$, and $3\pi/2$ between the background- and
$\pi\Delta$-subchannel amplitudes, respectively; in this model
initial- and final-state interactions are treated effectively and
allow for those relative phases. The access to interference effects
permit a cleaner separation of background and resonances. This in turn
makes it possible to make more reliable statements about the existence
and properties of nucleon resonances.

\begin{figure}[t!]
\centerline{\includegraphics[width=0.8\columnwidth]{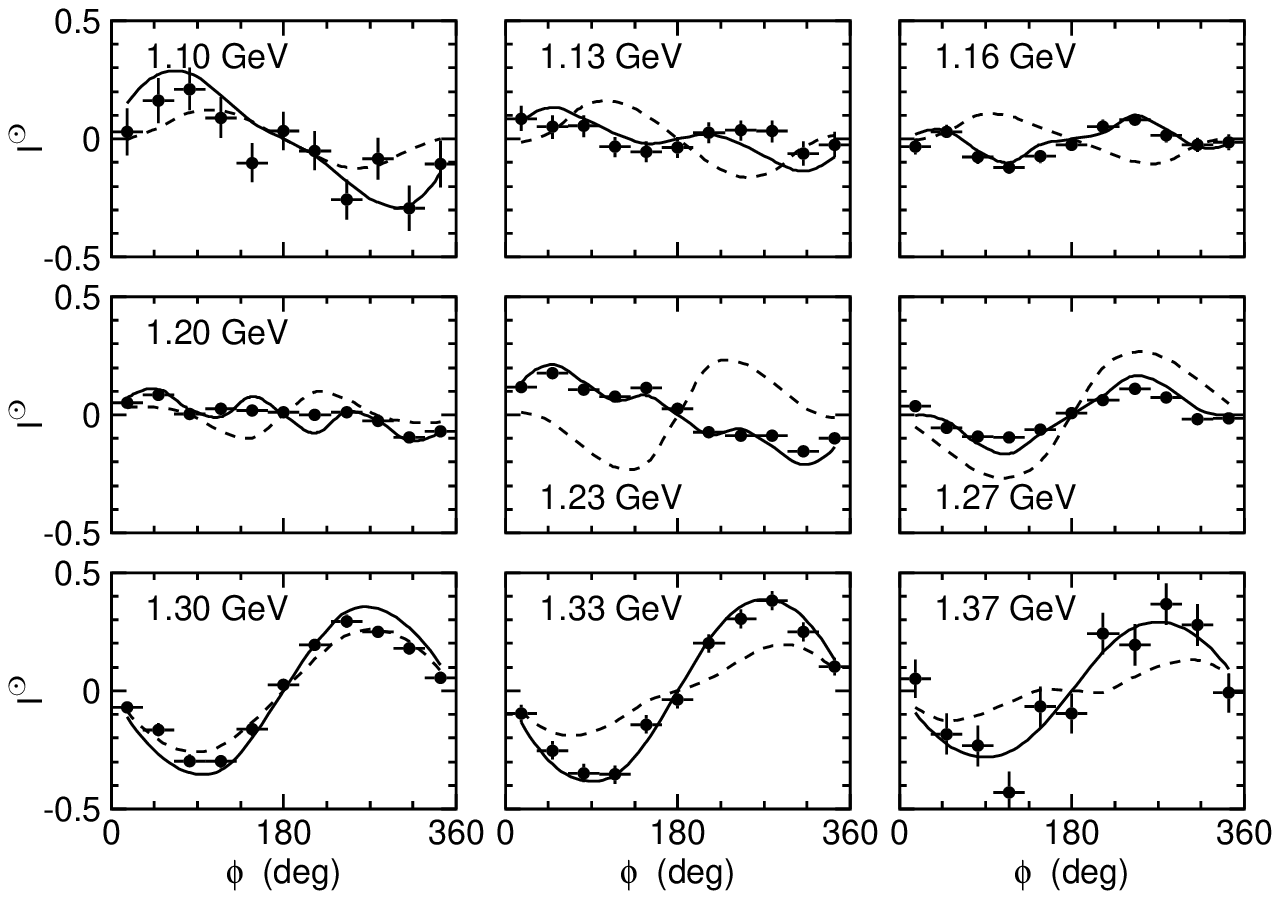}}
\caption{\label{fig:d13_a} Helicity asymmetries at $W=1.50$~GeV for
  nine bins of the invariant mass $M(p\pi^+)$ with its mean values
  indicated in each panel. The asymmetries are otherwise integrated
  over the full CLAS acceptance. The solid curves are the results of
  Mokeev {\it et al.};\protect\cite{Mokeev} the
  dashed curves show results of calculations by Fix and
  Arenh{\"o}vel.\protect\cite{Fix04}}
\vspace{3mm}
\centerline{\includegraphics[width=0.8\columnwidth]{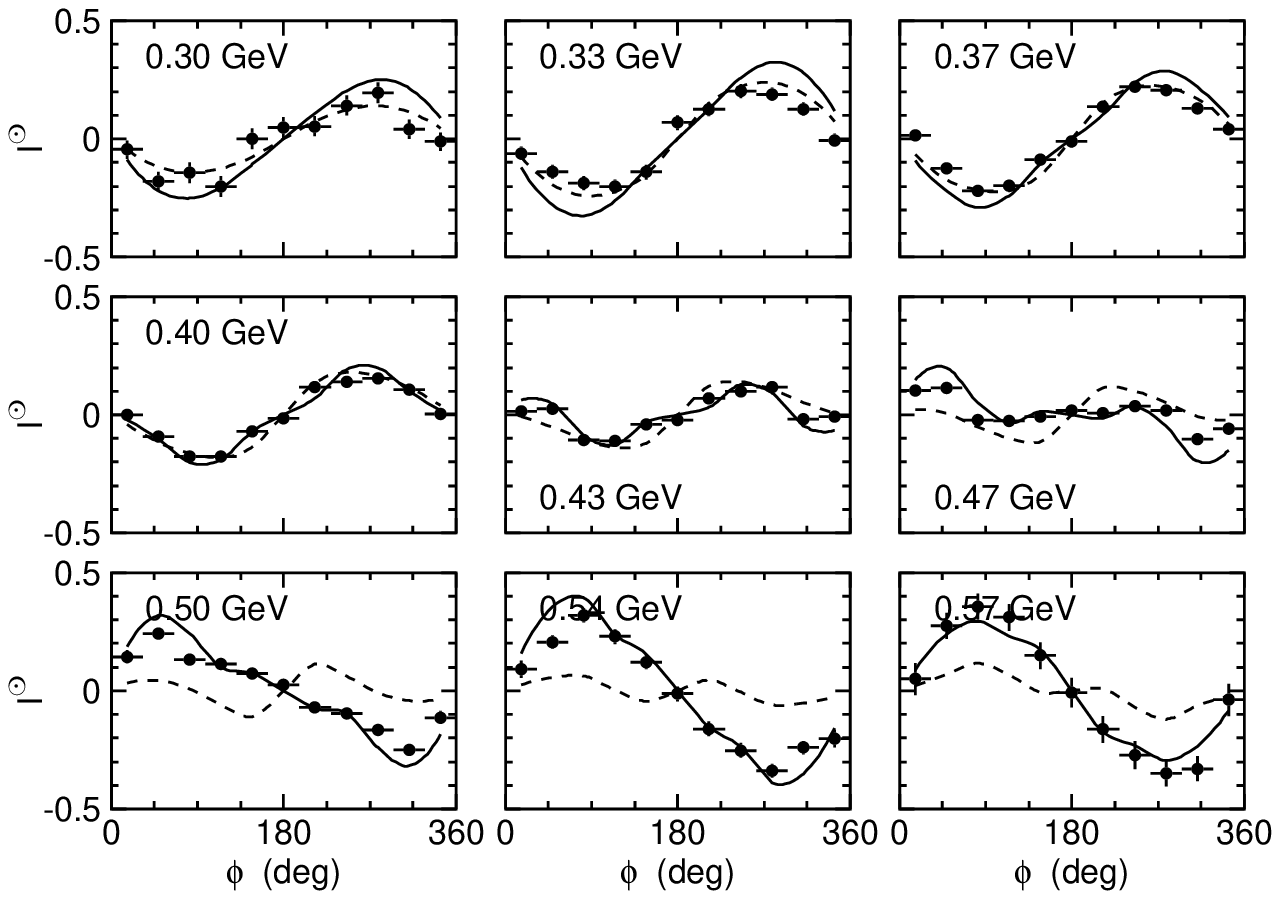}}
\caption{\label{fig:d13_b} Same as Fig.~\ref{fig:d13_a} for nine
  bins of the invariant mass $M(\pi^+\pi^-)$.}
\end{figure}

The large number of observed $\vec{\gamma}p\to p\pi^+\pi^-$ events
allows for a confident analysis of the data in selected kinematic
regions, making it possible to tune the different parts of the
production amplitude independently. Figures~\ref{fig:d13_a} and
\ref{fig:d13_b} give two examples of more selective distributions of
the same data around $W=1.50$~GeV binned into nine bins in the
invariant mass $M(p\pi^+)$ and into nine bins in the invariant mass
$M(\pi^+\pi^-)$, respectively.

\begin{figure}[ht!]
\centerline{
  \begin{minipage}[c]{0.55\linewidth}
    \includegraphics[width=\columnwidth]{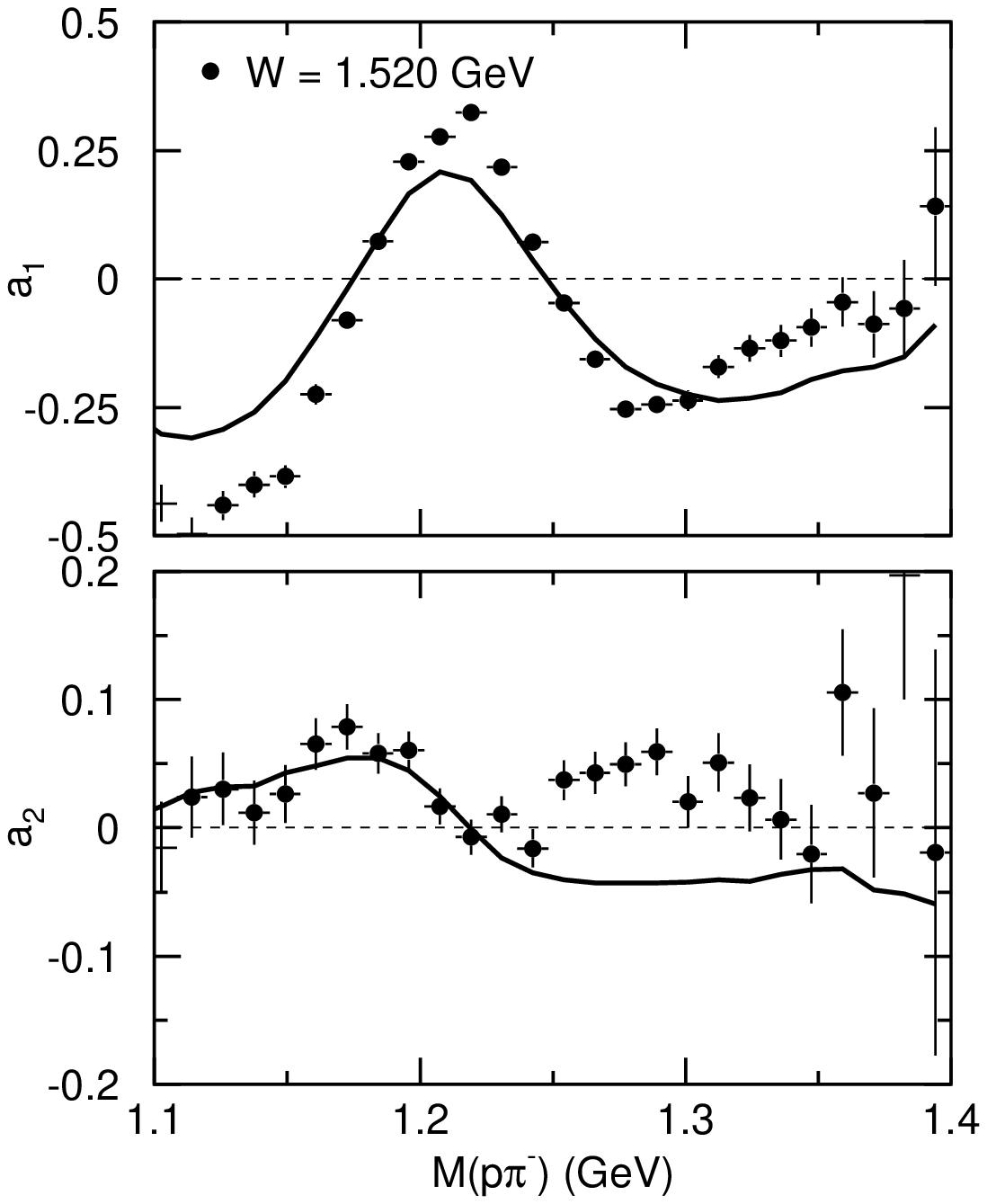}
  \end{minipage}\hfill
  \begin{minipage}[c]{0.35\linewidth}
    \includegraphics[width=\columnwidth]{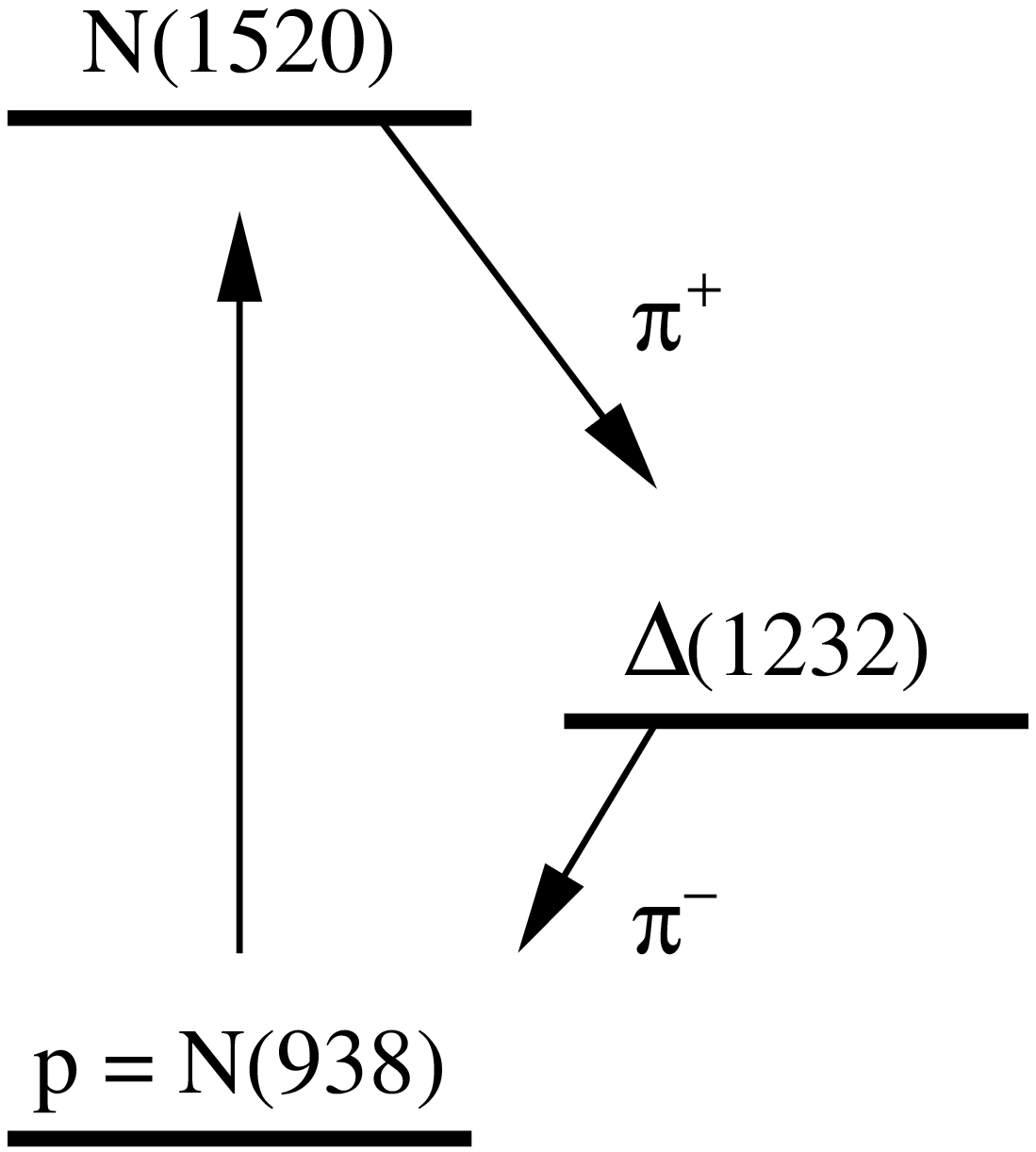}
  \end{minipage}
}
\caption{\label{fig:phicut}Fourier coefficients $a_1$ and $a_2$ as a
  function of the invariant mass $M(p\pi^-)$ for $W= 1.520$~GeV. The
  curves are the results of Fix and Arenh{\"o}vel.\protect\cite{Fix04}
  The vertical line indicates the masses of the $\Delta$ resonance.
  The diagram illustrates the sequential decay $N(1520) \to \pi^+ \Delta^0 \to \pi^+\pi^-p$
  which might be observed in the data.
}
\end{figure}

The double-pion final state allows further the study of sequential
decays of nucleon resonances such as $N(1520) \to \pi \Delta \to
\pi\pi N$. Figure~\ref{fig:phicut} shows the Fourier coefficients
$a_1$ and $a_2$ of the $\phi$ angular distributions as a function of
the invariant mass $M(p\pi^-)$ for $W = 1.520$~GeV.
The most interesting feature of these data is the change that occurs as
$M(p\pi^-)$ traverses the $\Delta(1232)$ resonance. A maximum is seen
in the region of this resonance. This hints at the way in which the
helicity asymmetry (along with other polarization observables) could
be used in studies of baryon spectroscopy.

\section{Outlook and Summary}

\begin{figure}[!ht]
\centerline{\includegraphics[width=0.8\columnwidth]{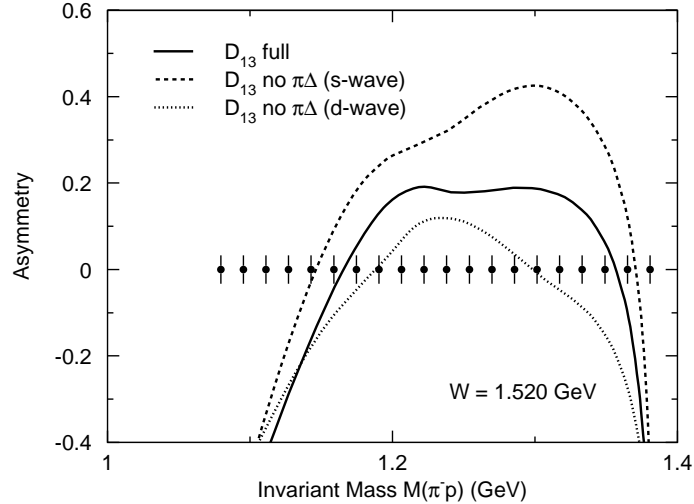}}
\caption{\label{fig:pz}Predictions of the double-polarization
  observable $P_z^\odot$ in the $\vec \gamma \vec p \rightarrow
  p\pi^+\pi^-$ reaction with circularly polarized photon beam and
  longitudinally polarized target at $W=1.520$ GeV.  The curves show
  results from a model by Fix and Arenh{\"o}vel\protect\cite{Fix04}
  for the full model (solid) and assuming no $s$-wave decay (dashed),
  or no $d$-wave decay (dotted) of the $N(1520)$ resonance. The points
  indicate the uncertainties which could be achieved in a proposed
  frozen-spin target experiment at CLAS\protect.\protect\cite{proposal05}}
\end{figure}

The observable $I^\odot$ discussed here is only one of many
polarization observables in the double-pion photoproduction
reaction. Many more polarization observables will be accessible in a
proposed experiment at Jefferson Lab using a transversely- as well as
longitudinally-polarized frozen-spin target.\cite{proposal05,Klein05} In
particular, we will be able to measure three single-polarization
observables ($P_x$, $P_y$, $P_z$) and nine double-polarization
observables ($P^{c,s}_x$, $P^{c,s}_y$, $P^{c,s}_z$, $P^\odot_x$,
$P^\odot_y$, and $P^\odot_z$) in the mass range up to 2~GeV/c$^2$.
Figure \ref{fig:pz} shows as an example predictions of the
double-polarization observable $P_z^\odot$ from the model of Fix and
Arenh{\"o}vel along with expected uncertainties of the proposed
data. The data will be able to differentiate between various
assumptions in the reaction dynamics; here, the question of the $s$-
and $d$-wave decay of the $N(1520)$ resonance into $\pi\Delta$.

Additional constraints for models or partial-wave analyses are
provided by recent studies of the $\gamma p \to \pi^0 \pi^0 p$
reaction channel, and are particularly expected from new
double-polarization experiments planned at ELSA.\cite{Thoma05}

In summary, we have given a brief overview of our \gp data, and we
have demonstrated, by means of phenomenological models, the
sensitivity of this helicity-asymmetry observable to the dynamics of
the reaction. Although existing phenomenological models do describe
unpolarized cross sections in the double-pion photo- and
electroproduction reaction, they have severe shortcomings in the
description of the beam-helicity asymmetries. In the region of
overlapping nucleon resonances (and model dependent backgrounds), it
clearly will be a challenge to any theoretical model to describe this
and other new observables from future experiments that depend so
sensitively on the interferences between them. Yet, without a proper
understanding of the $\pi\pi N$ channel the problem of the ``missing''
resonances is unlikely to be resolved.

\section{Acknowledgments}

This work was supported by the Italian Istituto Nazionale
di Fisica Nucleare, the French Centre National de la Recherche
Scientifique and Commissariat \`a l'Energie Atomique, the
U.S. Department of Energy and National Science Foundation, and the
Korea Science and Engineering Foundation. Southeastern Universities
Research Association (SURA) operates the Thomas Jefferson National
Accelerator Facility under U.S. Department of Energy contract
DE-AC05-84ER40150.


\begin{thebibliography}{50}

\bibitem{pdg04} S.~Eidelman {\it et al.}, {Phys.\ Lett.}
\textbf{B592}, 1 (2004).

\bibitem{capstick94} S.~Capstick and W.~Roberts, {Phys.~Rev.~D}
\textbf{49}, 4570 (1994).

\bibitem{pipiN-cs} ABBHHM Collaboration, {Phys.\ Rev.} \textbf{175},
  1669 (1968); ABBHHM Collaboration, {Phys.\ Rev.} \textbf{188}, 2060
  (1969); A.~Braghieri {\it et al.}, {Phys.\ Lett.}  \textbf{B363}, 46
  (1995); F.~H{\"a}rter {\it et al.}, {Phys.~Lett.}  \textbf{B401},
  229 (1997); M.~Wolf {\it et al.}, {Eur.~Phys.~J.} \textbf{A9}, 5
  (2000); Y.~Assafiri {\it et al.}, {Phys.\ Rev.\ Lett.}  \textbf{90},
  222001 (2003); M.~Ripani {\it et al.}, {Phys.\ Rev.\ Lett.}
  \textbf{91}, 022002 (2003); S.~A. Philips, Ph.D. thesis, The George
  Washington University (2003); M.~Bellis, Ph.D. thesis, Rensselaer
  Polytechnic Institute (2003); W.~Langg{\"a}rtner {\it et al.},
  {Phys.~Rev.~Lett.}  \textbf{87}, 052001 (2001).

\bibitem{pipiN-pol} J.~Ballam {\it et al.}, {Phys.\ Rev.\ D}
  \textbf{5}, 545 (1972); J.~Ahrens {\it et al.}  (GDH and A2
  Collaborations), {Phys.\ Lett.}  \textbf{B551}, 49 (2003).

\bibitem{Strauch05} S.~Strauch {\it et al.} (CLAS Collaboration),
  Phys.\ Rev.\ Lett. {\bf 95}, 162003 (2005).

\bibitem{Roberts04} W.~Roberts and T.~Oed, {Phys. Rev.} \textbf{C 71},
  055201 (2005); W.~Roberts, these Proceedings.

\bibitem{Mecking03} B.~A.~Mecking {\it et al.},
  {Nucl.~Instrum.~Methods} \textbf{A503}, 513 (2003).

\bibitem{Sober00} D.~I. Sober {\it et al.}, {Nucl.~Instrum.~Methods}
  \textbf{A440}, 263 (2000).

\bibitem{Maximon59} H.~Olsen and L.~C.~Maximon, {Phys.~Rev.~}
  \textbf{114}, 887 (1959).

\bibitem{Mokeev} V.~I.~Mokeev {\it et al.}, {Yad.\ Fiz.}  \textbf{64},
  1368 (2001), [Phys.\ At.\ Nucl. {\bf 64}, 1292 (2001)]; V.~Burkert
  {\it et al.}, {Nucl.~Phys.}  \textbf{A737}, S231 (2004);
  V.~I.~Mokeev {\it et al.}, Proc. NSTAR2004 (World Scientific, New
  Jersey, 2004), p. 321; V.~I.~Mokeev, these Proceedings.

\bibitem{Fix04} A.~Fix and H.~Arenh{\"o}vel, Eur.~Phys.~J. A {\bf 25},
  115 (2005).

\bibitem{Ukwatta05} T.N.~Ukwatta {\it et al.}, submitted to the
  Proceedings of the ``{\it VI Latin American Symposium on Nuclear
    Physics and Applications}'', Iguaz\'{u}, Argentina (2005).

\bibitem{Roca04} L.~Roca, {Nucl.~Phys.} \textbf{A748}, 192 (2005).

\bibitem{Strauch04} S.~Strauch {\it et al.}, Proc. NSTAR2004 (World
  Scientific, New Jersey, 2004), p. 317.

\bibitem{proposal05} Jefferson Lab proposal P06-013, ``{\it Measurement of
  $\pi^+\pi^-$ Photoproduction in Double-Polarization Experiments
  using CLAS}'', M. Bellis, V. Cred\'{e}, S. Strauch, spokespeople.

\bibitem{Klein05} F.~Klein, these Proceedings.

\bibitem{Thoma05} U.~Thoma, these Proceedings.
\end{thebibliography}
\end{document}